# Commuting Variability by Wage Groups in Baton Rouge 1990-2010[1]

Yujie Hu[1], Fahui Wang[1]*, Chester G Wilmot[2]

[1]Department of Geography and Anthropology, Louisiana State University, Baton Rouge, LA 70803, USA
[2]Department of Civil and Environmental Engineering, Louisiana State University, Baton Rouge, LA 70803, USA

**Abstract**
Residential segregation recently has shifted to more class/income-based in the United States, and neighborhoods are undergoing significant changes such as commuting patterns over time. To better understand the commuting inequality across neighborhoods of different income levels, this research analyzes commuting variability (in both distance and time) across wage groups as well as stability over time using the CTPP data 1990-2010 in Baton Rouge. In comparison to previous work, commuting distance is estimated more accurately by Monte Carlo simulation of individual trips to mitigate aggregation error and scale effect. The results based on neighborhood's mean wage rate indicate that commuting behaviors vary across areas of different wage rates and such variability is captured by a convex shape. Affluent neighborhoods tended to commute more but highest-wage neighborhoods retreated for less commuting. This trend remains relatively stable over time despite an overall transportation improvement in general. A complementary analysis based on the distribution of wage groups is conducted to gain more detailed insights and uncovers the lasting poor mobility (e.g., fewer location and transport options) of the lowest-wage workers in 1990-2010.

## Introduction

Journey-to-work trips occur on a daily basis by multiple transportation modes (e.g., drove-alone, carpool, public transit, bicycle and walk). Even though commuting represents only 20-25% share of all-purpose trips in the United States (Sultana 2002; Horner 2004), it results in two of the most congested periods in a day and establishes the major transportation infrastructure and service needs; therefore, it is strongly connected with major public policy issues such as traffic congestion, air pollution and greenhouse gas emissions (Sultana and Weber 2014). As a trip linking home (residential areas) to employment (commercial, industrial and other land uses), commuting is related with land use patterns (AASHTO 2013). For example, as population and jobs become increasingly decentralized, commuting distances are reported to become lengthier, while times remain stable or even dropped over time (Gordon et al. 2004; Hu and Wang 2016). Yet, such commuting patterns were found to vary for workers of different sociodemographic groups such as income (Shen 2000).

Recently, income is reported to be another major factor that affects residential segregation besides race-ethnicity in the United States (Massey et al. 2009; Niedzielski et al. 2015). For instance, the number of high-poverty neighborhoods is increasing and they are spreading from central cities to suburban areas (Cooke and Marchant 2006; Kneebone and Garr 2010). In addition, structures of jobs are also changing; for example, the number of low-skilled jobs (usually less-payed) is growing in many suburban locations and across metropolitan areas in the United States (Niedzielski et al. 2015). Besides the increasing role in altering residential and employment layouts, there are several other reasons for this research to target income. Specifically, income plays a determining role in a household's residential choice that is usually driven by the tradeoff between commuting length and housing size as suggested by the urban economic model; it is perhaps the most important determinant in vehicle ownership and thus affects mobility (commuting); and finally it is freely available from Census (e.g., CTPP) at the neighborhood (e.g., census tract) level, which reflects the neighborhood attributes and provides the geographic context for urban research. Therefore, more research is needed to study the role of income in affecting commute patterns as well as the (in)stability of the effect over time. Lack of access to data of individual commuters, particularly quality data over a long time period, prevents us from addressing the question directly. Nevertheless, analysis of commuting variability by neighborhood income levels may still shed light on the issue (Wang 2003).

This article compares and analyzes commute distance and time for different neighborhood income groups (wage as a surrogate) for three time periods. The study is based on Baton Rouge with data extracted from the 1990-2010 Census Transportation Planning Package (CTPP). Several hypotheses are examined: (1) workers with higher wage rates tend to commute more, (2) the trend may differ by commuting time and distance, and (3) the trend described in (1) may have altered over time.

**Commuting variability by income and commuting length**
As commuting is the trip connecting home sites to jobs, it is largely affected by the spatial separation between a worker's residence and employment locations. Therefore, we can analyze workers' commuting variability by examining where they are and who they are (Wang 2001). For example, some studies explain commuting by spatial factors such as land use patterns (Wang 2000; Sultana 2002; Wang 2003; Horner 2004; Horner 2007; Hu and Wang 2016). On the other hand, workers of various socio-demographic characteristics in the same neighborhood may respond differently in commuting choice. Many studies emphasize aspatial factors such as race (Kain 2004), wage (Wang 2003), income (Horner and Schleith 2012), and gender (Kwan and Kotsev 2015). This article specifically focuses on the aspatial factors.

Existing studies commonly rely on aggregated socioeconomic data such as from Census, which would make such analyses vulnerable and open to criticism due to the ecological nature, however. This research focuses on only one specific aspatial factor to examine its connection with commuting, instead of investigating all above factors together. As mentioned, we are particularly interested in understanding the relationship between commuting and neighborhood income, which is driven by the recent change of residential segregation in the United States from racial-ethnic to more class/income-based (Massey et al. 2009; Niedzielski et al. 2015).

Much research has already recognized the determining role of income in workers' housing (thus commuting) choice (e.g., Cervero et al. 2006). For example, low-wage workers are reported to spend a much higher proportion of their income on commuting (6.1%) than other workers (3.8%); furthermore, the working poor who rent spend a greater portion of their income on combined costs of commuting and housing (32.4%) than other workers (19.7%) (Roberto 2008). Low-wage workers also have significantly lower vehicle ownership than others (Lowe and Marmol 2013), and are thus more likely to use slower transportation modes such as public transit, carpool, bicycle and walk than their high-wage counterparts (Ross and Svajlenka 2012; McKenzie 2014). To this end, higher-wage workers may generally commute longer than lower-wage workers because of their better mobility and economic conditions. To study the link between commute and income, there are generally two ways. The first type of research is to develop analysis based on actual commuting flow data, for example, Census Transportation Planning Package (CTPP) and Longitudinal Employer-Household Dynamics (LEHD) provided by the U.S. Census. CTPP delivers commute flow data for general workers or a few sets of single classes such as poverty status, household income and transportation mode. It is summarized at multiple geographic scales (e.g., census tracts and traffic analysis zones) and temporal resolutions (e.g., 10-year, 5-year, and 3-year census) from 1990 to now. Based on 1990 CTPP, Wang (2003) measured commuting lengths (in both distance and time) for neighborhoods of different wage levels in Cleveland and found that compared to time, commuting distance was more sensitive to wage in a way that wealthier neighborhoods overall commuted longer than lower-wage neighborhoods, but the wealthiest wage group shortened their commute slightly. With more detailed spatially and socially disaggregated data than CTPP, the LEHD data provides actual observed commute flows such as by individual income groups at census block level. For example, Horner and Schleith (2012) investigated the commuting pattern by three income groups based on 2002-2010 LEHD for Leon County, Florida, and found that average observed and minimum commute distances got lengthier as income increased. Based on LEHD 2010, a recent study by Niedzielski et al. (2015) analyzed commuting variability by three income groups for a medium size city Wichita, Kansas, and reported that high-income groups commuted significantly longer distances than low-income groups. Another body of research uses spatial interaction models to synthesize commute flows by specific subgroups when such information is not available. One typical approach refers to the information minimization technique designed by O'Kelly and Lee (2005). Their approach has been widely used to synthesize disaggregated commute trips, such as trips by a single factor like occupation (O'Kelly and Lee 2005) and race/ethnicity (Jang and Yao 2014), or by any combined factors like gender-occupation (Sang et al. 2011) and race-income (Niedzielski et al. 2015). The effectiveness of this technique is validated by a recent research that compared synthetic commute flows by income group with the actual flows by income group extracted from LEHD for Wichita, Kansas (Niedzielski et al. 2015).

The issue of commuting metric is another key point to this research. Most studies use commute time to measure commute length due to its wide availability from survey data (e.g., CTPP in the United States). Some argue that commute distance is a better metric (Van Ommeren and Gutiérrez-i-Puigarnau 2011) since mileage could provide a more consistent measure of commuting length (Sultana and Weber 2007). Among the few studies on commute distance, most use a zonal centroid-to-centroid approach, including those based on Euclidean distance (Gera 1979; Hamilton 1982; Levinson and Kumar 1994; Levinson 1998; Horner and Murray 2002; Clark et al. 2003; Wang 2003; Kim 2008) and others measuring in network distance (White 1988; Levinson and Kumar 1994; Cervero and Wu 1998; Wang 2000; Wang 2001; Horner 2002; Yang 2008). Either approach may underestimate actual distances, as it still assumes that all people start and end a journey at the zonal centroids. Hewko et al. (2002) noted aggregation error resulting from using a centroid to represent a neighborhood in distance measure would be significant for analyses using more aggregated units such as census tracts. The measurements of these commute metrics are also subject to the modifiable areal unit problem (MAUP) (Niedzielski et al. 2013). A more accurate measure for commute length is needed in commuting studies, especially when aggregated data such as CTPP and LEHD are used (Hu and Wang 2015a).

Even though LEHD data is distributed at finer spatial and social levels than CTPP (e.g., census block vs. census tract, neighborhood wage rate vs. individual income), it is limited to three income groups and to a short time period only (i.e., 2002-2014). Besides, LEHD data does not report commute lengths (e.g., times) as CTPP does. Furthermore, the utility of the technique to synthesize commute flows is by far limited to Wichita, Kansas only and still unknown to other areas. As a result, we employed CTPP 1990-2010 data at census tract level to examine the commuting length variability (both distance and time) across neighborhood income groups as well as its (in)stability of the pattern over time in a medium-sized city Baton Rouge, Louisiana. Specifically, we use wage rate earned by resident workers instead of household income in this study because of better data from the CTPP on wage, and also for the closer tie of wage to commuting behavior than other income sources (Gera and Kuhn 1980; Wang 2003). Nevertheless, wage and income are two closely related measures. In comparison to previous work, commuting distance is estimated more accurately by Monte Carlo simulation of individual trips to mitigate the aforementioned aggregation error and scale effect; and it is further advanced by integrating land use pattern in the simulation process to improve the accuracy.

**Study Area and Data Sources**

East Baton Rouge Parish (i.e., county) in Louisiana, the core of Baton Rouge metropolitan area, is selected as the study area (see Figure 1). As one of the fastest growing metropolitan areas in the South, the commuting pattern in East Baton Rouge Parish may change significantly over time, which makes it an ideal study area for this research. In addition, most part of this Parish is the city of Baton Rouge, a medium-sized city. Looking into the commuting pattern and its (in)stability could provide a unique case study in contrast to large cities used in most commuting studies. This parish has a total population of 440,000 (there are 182,705 resident workers who reside and also work in this parish) with an area of 471 square miles in 2010. For simplicity, hereafter the study area is referred to as Baton Rouge.

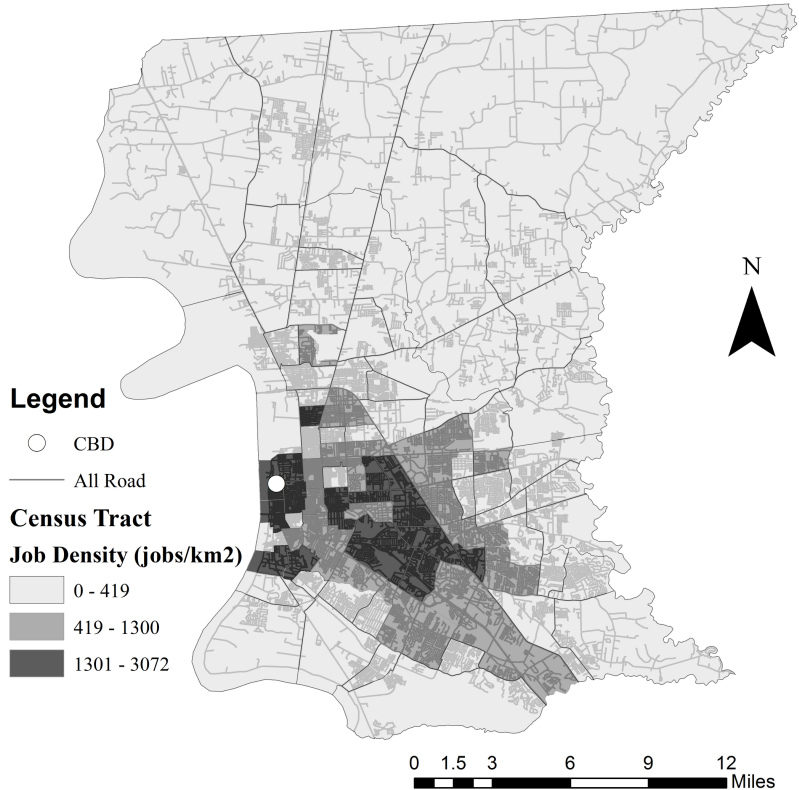

Figure 1. Baton Rouge in 2010

Major data sources are the Census Transportation Planning Packages (CTPP): the 1990 and 2000 CTPPs extracted from the long form decennial census with a sampling rate of about one in six households (www.transtats.bts.gov), and the most recent 2006-2010 CTPP from the 5-year American Community Survey, a short form survey sampled about one in forty households (ctpp.transportation.org/Pages/5-Year-Data.aspx). Note that the 2006-2010 data and corresponding analyses are hereafter referred to as 2010 for simplicity. All CTPPs consist of three parts: Part 1 on residential places (e.g., number of resident workers and breakdowns by wage range, and mean wage rate in each zone), Part 2 on workplaces (e.g., number of jobs in each zone), and Part 3 on journey-to-work flow (e.g., number of commuters from a residence zone to a workplace zone, and average commute time and breakdowns by different transportation modes).

There is an inconsistency in area units used in the CTPP data for Baton Rouge: only traffic analysis zones (TAZs) were used in 1990, multiple zonal levels were used in 2000 (census tracts, census block groups, and TAZs for Parts 1 and 2, only census tracts for Part 3), and census tracts and TAZs were used in 2010. We chose census tracts as the unit to use throughout. In Baton Rouge, the 1990 TAZs were mostly components of census tracts for easy aggregation with only very few minor exceptions. There were 85, 89 and 91 census tracts in Baton Rouge in 1990, 2000 and 2010, respectively (excluding the 2010 airport tract where no records of any resident workers or jobs are provided in the data). The slightly increased number of tracts in later years were simply the result of split tracts from earlier years (i.e., 2000 vs. 1990, 2010 vs. 2000). This enabled us to integrate the data in three time epochs based on the 85 census tracts in 1990 when needed. Corresponding spatial data sets in GIS (including census tracts, TAZs and road networks) were extracted from the TIGER Products 1994, 2000, and 2010 from the U.S. Census Bureau. We are aware of the discrepancy in the samples between long form CTPP and short form CTPP as well as the time gaps of using the 1994 and 2010 GIS data to match the 1990 and 2006-2010 CTPP, respectively, but they were the best data accessible to us.

The National Land Cover Database (NLCD, http://www.mrlc.gov), a national land cover product, is used to help improve the accuracy of individual trip simulation, in particular the simulation of trip destinations. Three NLCD products—e.g., NLCD 1992, 2001, and 2011—are employed to match with the above CTPP and TIGER data. The NLCD has a spatial resolution of 30 square meters, and only one land cover type is recorded in each pixel, 30m × 30m polygon. It provides a uniform land cover classification across the entire United States, and is perhaps the most accessible and commonly used national land cover map (Jin et al. 2013). In this article, the high intensity developed areas that are commonly interpreted as commercial/industrial lands are used to define the geographic areas for simulated job locations. The geographic areas of resident workers, however, are calibrated on the basis of census block population data due to its better accuracy in capturing the residential pattern than the NLCD. Table 1 summarizes the data sources used in this research.

Table 1. Summary of all data sources

| Data layer | Year | Spatial scale | Format | Source | Purpose |
|---|---|---|---|---|---|
| CTPP | 1990 | TAZ | ASCII/excel file | BTS | Total number of workers, jobs, and commuters |
| | 2000 | Census tract | ASCII/excel file | BTS | |
| | 2010 | Census tract | ASCII/excel file | AASHTO | |
| Zone boundary | 1994 | TAZ | Vector/shapefile | TIGER/Line | Define boundary of zone units |
| | 2000 | Census tract | Vector/shapefile | TIGER/Line | |

|  | 2010 | Census tract | Vector/shapefile | TIGER/Line |  |
| --- | --- | --- | --- | --- | --- |
| Population | 1990 | Census block | ASCII/excel file | Census | Spatial extent of residential areas |
|  | 2000 | Census block | ASCII/excel file | Census |  |
|  | 2010 | Census block | ASCII/excel file | Census |  |
| NLCD | 1992 | 30m×30m cell | Raster/tif file | MRLC | Spatial extent of workplaces |
|  | 2001 | 30m×30m cell | Raster/tif file | MRLC |  |
|  | 2011 | 30m×30m cell | Raster/tif file | MRLC |  |
| Road network | 1994 | - | Vector/shapefile | TIGER/Line | Define entire road network |
|  | 2000 | - | Vector/shapefile | TIGER/Line |  |
|  | 2010 | - | Vector/shapefile | TIGER/Line |  |

## Measuring Commuting Distance and Time
### Calibrating Commuting Distance

Unlike commute time, distance is not reported in the CTPP and thus needs to be estimated. As mentioned above, the traditional centroid-to-centroid distance approach may return substantially biased estimates. Given that, Hu and Wang (2015a; 2015b) designed a Monte Carlo simulation approach to obtain journey-to-work trips between individual points, and showed great improvements over the centroid approach. Specifically, the approach generates corresponding numbers of resident workers and jobs in each zone, following their frequency distributions. In terms of the locations of resident workers/jobs, it is assumed that they are located completely randomly in an entire zone. By doing so, it improves over the centroid approach by spreading the simulated trip origins/destinations across an area instead of concentrating at a single centroid, but also raises other concerns. For example, when relying on the data at census tract level, it may generate resident workers/jobs in water or forest areas and does not account for varying densities within a tract.

This article further advances the Monte Carlo simulation approach by considering land use patterns in the simulation process. Specifically, we simulate the locations of resident workers only in places of residential land use based on census block data, and jobs only in places of commercial/industrial land use as determined using the NLCD. For example, Figure 2 illustrates the distribution of commercial/industrial land use patterns in Baton Rouge 2010. Compared to the previous approach in Hu and Wang (2015a; 2015b) that randomly generated job places within the entire census tracts, this article limits the potential locations of jobs to only the commercial/industrial land use areas as marked by black polygons in Figure 2. Significant differences between the two methods are observed in outskirt areas (e.g., the northeastern region), where census tracts of large area sizes have far fewer job locations. By accounting for the spatial variability in land use type and intensity, this improvement simulates the locations of resident workers and jobs that are more consistent with their actual spatial patterns, and leads to more accurate distance measures. See Figure 3 for an illustration.

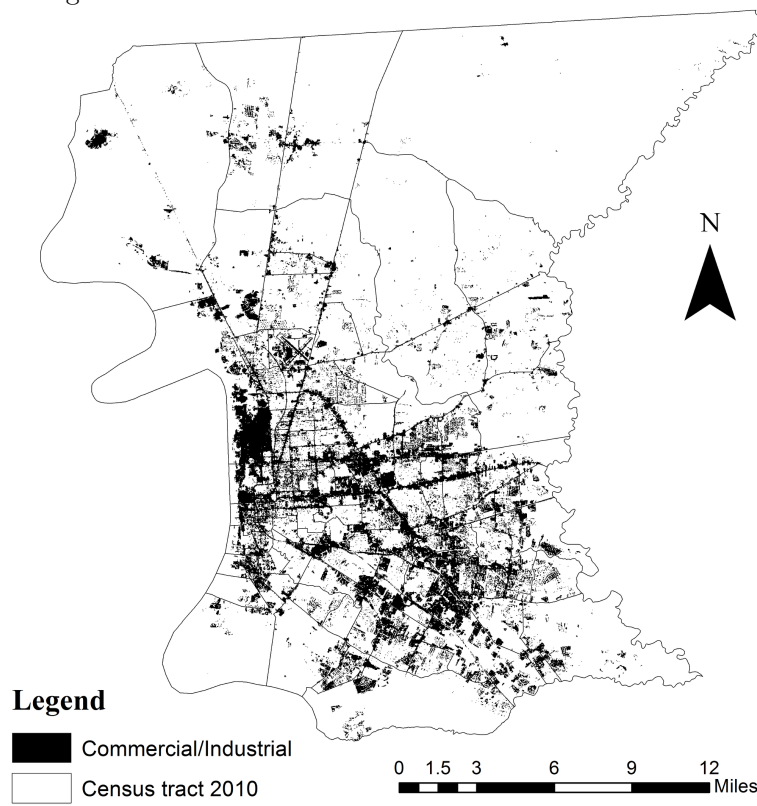

Figure 2. Commercial/Industrial land use in Baton Rouge 2010

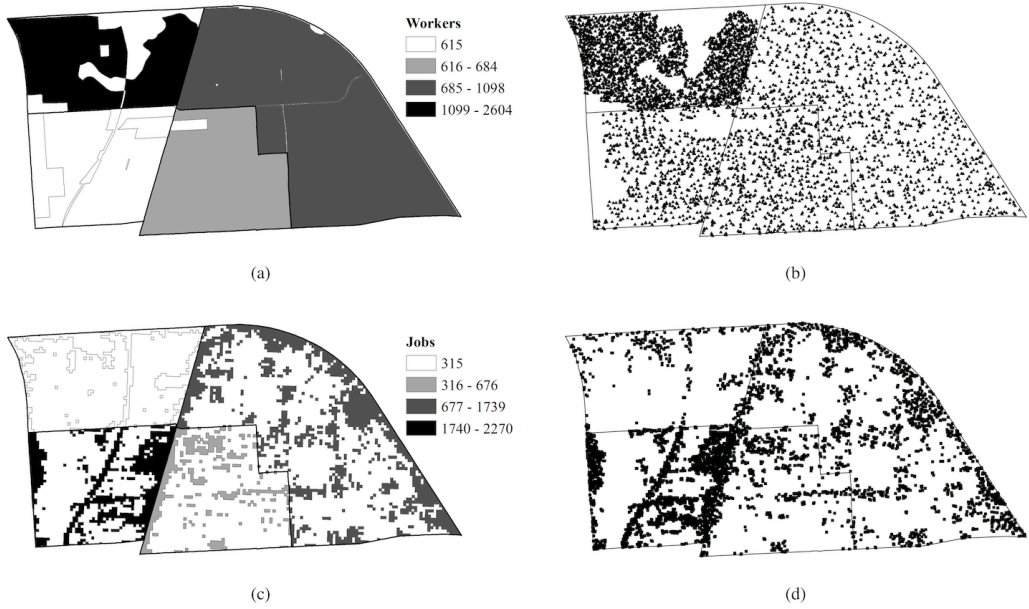

Figure 3. Spatial distribution of (a) resident workers in residential land use areas in zones, (b) simulated resident workers, (c) jobs in commercial or industrial land use areas in zones, and (d) simulated jobs

Briefly, the first task of our approach is to randomly generate points of resident workers and jobs in corresponding land use areas in tracts so that their total numbers at the tract level are proportional to the observed patterns of resident workers and jobs in the CTPP. Specifically, denoting the total numbers of simulated and actual commuters by n and N, respectively, and the numbers of resident workers and jobs in a tract i reported from the CTPP by $R_i$ and $E_i$, the numbers of resident workers and jobs to be simulated in the tract i in this step are $(n/N)R_i$ and $(n/N)E_i$, respectively (Hu and Wang 2015b). The second task is to pair the origins (workers) and destinations (jobs) to form OD trips that follow a discrete frequency distribution that is also consistent with the reported journey-to-work flows in the CTPP, and then measure the network distance for each OD trip. In other words, randomly select a resident worker and a job simulated in the previous step from tract i and j, respectively, and then match them together to form a trip. The trajectories of each trip are retrieved by shortest network distance. Denote the actual commuting flow from tract i to tract j is $x_{ij}$, the simulated trips when aggregated at the tract level is $(n/N)x_{ij}$, i.e., proportional to the actual journey-to-work flow pattern (Hu and Wang 2015b). In our analysis, n is set to be equal to N, i.e., 182,705.

It would be desirable to take mode choices into consideration when measuring travel distance for each simulated journey-to-work trip since people commute by multiple modes. For example, Wang (2000) recovered commute distance based on the centroid-to-centroid network distance in Chicago by two major modes, i.e., vehicles (including drove-alone, carpool, bus, taxi, etc.) and trains (subway and rail), due to the high percentages of commuters in both modes. In Baton Rouge, however, the majority commuted by auto (including drove-alone and carpool), and the percentage was steady over time (i.e., 94-95%; Hu and Wang 2016). Therefore, we estimated commute distance solely for auto travel without considering other modes.

The simulation approach disaggregates the reported zonal journey-to-work flow into individual trips, and permits more accurate estimation of commute distances by mitigating the aggregation error and zonal effect. For example, the mean within-tract commute distance for the most northeast tract in 2010 is 0 by the centroid-to-centroid measure, and becomes 8.39 miles by the simulation approach without considering land use patterns, and 7.74 miles by the simulation method accounting for land use variability. In this case, the land use-integrated simulation approach returns a relatively smaller intrazonal commute distance with comparison to the basic simulation approach without considering land use patterns as areas with possible resident workers/jobs are closer to each other than what the random pattern suggests.

**Measuring Mean Commuting Distance and Time**

The mean commuting distance/time of a zone is a common measure of commuting pattern (Gera 1979; Gera and Kuhn 1980; Gordon et al. 1989a; Gordon et al. 1991; Giuliano and Small 1993; Cervero and Wu 1998; Wang 2000; Wang 2001; Wang 2003; Kim 2008; O'Kelly et al. 2012; Sultana and Weber 2014), and is defined as the average travel distance/time from one zone to all zones weighted by corresponding number of commuters.

$$MC_i = \sum_{j=1}^{n}\left(\frac{f_{ij}c_{ij}}{R_i}\right) \qquad (1)$$

As formulated in Equation 1, $MC_i$ is the mean commute in tract i; $f_{ij}$ is the commuter flow residing in tract i and working in tract j; $c_{ij}$ is travel cost (i.e., distance/time) between tract i and j; $R_i$ is the number of resident workers in tract i; and n is the total number of tracts in the study area. It indicates how far/long resident workers in tract i, on average, commute. We measured mean commute time in this manner by using the journey-to-work data from the CTPP Part 3 such as the reported commuter flow $f_{ij}$ and average commute time $c_{ij}$ (all travel modes).

$$MC_i = \sum_{k=1}^{m}\left(\frac{f_k c_k}{R_i}\right), \qquad (2)$$

where $f_k = 1$ if trip k starts in tract i and ends in tract j, 0 otherwise.

Mean commute distance is calibrated by the above simulation approach.

Equation 1 is now revised as Equation 2 for simulated trips. Specifically, if a simulated trip k starts in tract i and ends in tract j, $f_k$ then equals 1, indicating one eligible trip; $c_k$ is the network distance measured for trip k; m = 182,705, denoting the total number of simulated trips.

**Commuting Patterns by Neighborhood's Average Wage**

People of different wage rates may have varying responses to the classic tradeoff between commute length and house size. Without access to data of individual commuters, this section investigates the effect of wage on commuting by using a neighborhood's mean wage rate extracted from the CTPP Part 1.

To account for the effect of wage inflation over time, we group the tracts in each year by their mean wage rate percentiles such as 20, 40, and so on, as shown in Table 2. In 1990 and 2000, the tract-level mean commuting distance increased with the mean wage rate up to about the middle point (i.e., 40-60 percentile) and then declined toward the wealthiest neighborhood. This is largely consistent with the finding of Cleveland in 1990 reported in Wang (2003). The average commute distance peaked at the middle-range wage neighborhoods in our study area but more to the side of upper-middle wage neighborhoods in Cleveland. In other words, the response of mean commute distance to rising mean wage in neighborhoods may be characterized by a convex shape in 1990 and 2000. This is further confirmed by the convex curves in Figure 4 (for 1990 as an example) and the regression results in Table 3 (note the + and - signs for the coefficients of mean wage and its square terms, respectively, in 1990 and 2000, and both are statistically significant in either year). However, this pattern was less clear and not significant in 2010 because the middle 40-60-percentile tracts experienced a minor dip in commute distance; and in general, higher wage groups travelled farther than lower wage groups (e.g., 7.14 and 6.43 miles by the 60-80 and 80-100-percentile groups, respectively, vs. 5.40 and 6.33 miles by the 0-20 and 20-40-percentile groups, respectively).

Table 2. Mean commute distance and time by tract-level mean wage rate

| Mean Wage Percentile | Wage cutoff point | | | Mean commute distance (mile) | | | Mean commute time (min) | | |
|---|---|---|---|---|---|---|---|---|---|
| | 1990 | 2000 | 2010 | 1990 | 2000 | 2010 | 1990 | 2000 | 2010 |
| 0-20 | 14,822 | 21,085 | 27,529 | 4.58 | 5.10 | 5.40 | 16.78 | 20.39 | 17.51 |
| 20-40 | 18,420 | 24,140 | 34,905 | 4.68 | 5.67 | 6.33 | 16.50 | 19.40 | 19.32 |
| 40-60 | 22,766 | 31,915 | 45,023 | 7.95 | 7.53 | 5.96 | 18.66 | 19.38 | 17.52 |
| 60-80 | 28,050 | 41,680 | 53,551 | 6.72 | 7.28 | 7.14 | 16.57 | 18.83 | 18.10 |
| 80-100 | 42,760 | 63,245 | 75,899 | 5.84 | 5.21 | 6.43 | 15.13 | 15.49 | 17.50 |

| Total | - | - | - | 5.95 | 6.17 | 6.25 | 16.73 | 18.73 | 17.98 |

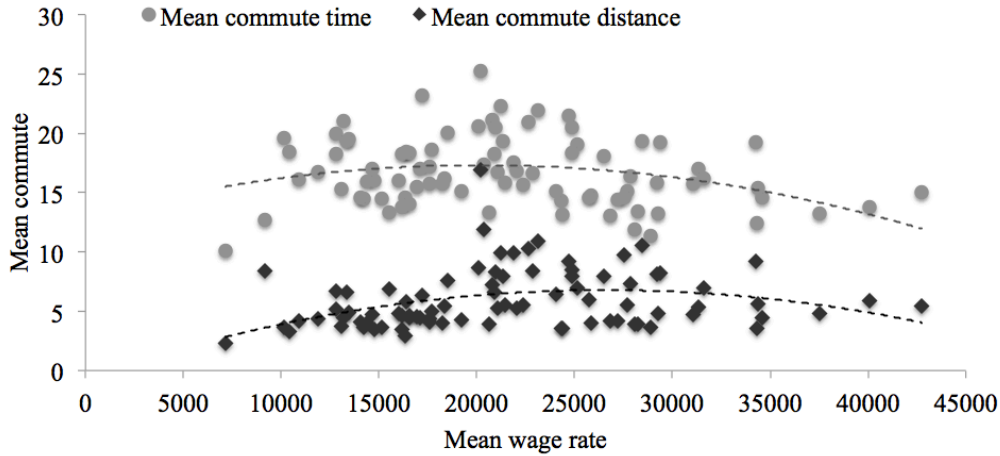

Figure 4. Mean commuting distance and time vs. a tract's mean wage rate in 1990

Table 3. Regression of mean commute vs. mean wage rate during 1990-2010

| Dependent variable | Mean commute distance | | | Mean commute time | | |
|---|---|---|---|---|---|---|
|  | 1990 | 2000 | 2010 | 1990 | 2000 | 2010 |
| Intercept | -0.61 | -0.19 | 3.12 | 13.02*** | 16.97*** | 13.21*** |
|  | (-0.31) | (-0.10) | (1.63) | (5.52) | (6.02) | (4.91) |
| Mean_wage | 0.001** | 0.0004*** | 0.0001 | 0.0004* | 0.0002 | 0.0003 |
|  | (3.16) | (3.52) | (1.30) | (2.03) | (1.42) | (1.91) |
| Mean_wage$^2$ | -1.0E-8** | -5.9E-9*** | -1.1E-9 | -1.1E-8* | -5.5E-9* | -3.2E-9 |
|  | (-2.85) | (-3.55) | (-0.95) | (-2.41) | (-2.22) | (-1.95) |
| F value | 5.84 | 6.32 | 2.58 | 4.30 | 11.43 | 1.91 |
| $R^2$ | 0.13 | 0.13 | 0.06 | 0.10 | 0.21 | 0.04 |
| No. observations | 85 | 89 | 91 | 85 | 89 | 91 |

Note: *t*-statistics are in parentheses; *significant at the 0.05 level, **at the 0.01 level, ***significant at the 0.001 level (two-tailed test).

The convex shape pattern for mean commute distance may first reflect the complex residential choice behavior in the tradeoff between commuting cost and house space. As the income net of commuting cost is larger for higher-wage workers and their housing expenditures increase with income, they opt to live farther away from the central city (and their jobs in general) for more house space and better community environment (Gordon et al. 1989b; Kim 2008). In addition, the individual's job location behavior would induce workers to locate their jobs at a distance from their residential places in order to achieve maximization in their earnings net of travel cost and job-searching spending. The job location behavior is more usual in a job-decentralized city (Gera and Kuhn 1980). The third possible reason refers to the nature of their jobs. Specifically, higher-wage workers usually have jobs requiring more specialties, and thus may need to commute further for appropriate jobs; on the contrary, lower-wage workers usually take less skilled jobs and may find suitable employment everywhere (Prashker et al. 2008). All explanations rely on better transport mobility for workers in higher-wage neighborhoods. Mobility for the workers in lower-wage neighborhoods is more limited (e.g., low vehicle ownership, high dependency on bicycling, walking, or transit that is often feasible only in the central city area). However, this increasing trend of commute distance may be reversed when the mean wage rate reaches a certain level. A higher wage rate also means a higher opportunity cost for more commuting. More importantly, workers living in the neighborhoods of highest wages can also afford homes that are closer to jobs and may command a higher unit price of housing. In short, the high wage enables the workers of this group to cut back on their commute lengths without sacrificing house space.

In terms of mean commute time, the convex shape pattern remained valid in 1990 (i.e., mean commute time peaked at 18.66 minutes in the tracts of the 40-60-percentile wage group), but did not hold in 2000 or 2010. In 2000, clearly the mean commute time was the highest for the poorest tracts and declined steadily to wealthier tracts. In 2010, the highest mean commute time was experienced by the tracts of 20-40-percentile wage group, and varied within a narrow band (i.e., 17.5-18.1 minutes) in the tracts of the other wage groups.

The largely inconsistent trends between the mean commute distance and time were attributable to the variability of mode distributions across tracts of various mean wage rates, as distance was measured by network distance while time was extracted from the reported all mode-based travel time in CTPP. It is well-known that travel time is longer for carpool than drove-alone, and even longer for public transits and others. As shown in Table 4, the percentage of commuters by drove-alone tracts was the smallest in the neighborhood of the lowest wage rate (i.e., 61.9% in 1990, 62.2% in 2000 and 66.5% in 2010), and increased gradually to tracts in higher wage rate. In other words, more workers had to use slower transport modes in lower-wage neighborhoods, and thus increased their commute time. Furthermore, as more low-wage workers tended to live in central city with higher densities and more congested roads, even those drove-alone commuters would use more time to travel the same distance and were less likely to convert their shorter distance trips to shorter duration.

As discussed previously, a shorter commute distance also reflects less mobility in job search. The double disadvantages were most evident in 2000 when the tracts in the 0-20-percentile wage group spent most time to commute the shortest distance. As Niedzielski and Boschmann (2014) noted, the relationship between travel distance and time is not always monotonic (e.g., spending less time to travel shorter distance); as a result of the diverse socioeconomic attributes of commuters, in fact, we are more likely to observe such a nonmonotonic pattern between commute distance and time for different commuter groups from the same area as what we discovered in Table 2.

Table 4. Commuting modal splits by tract-level mean wage rate

| Year | Wage percentile group | Drove-alone | Carpool | Public transit | Others[1] |
|---|---|---|---|---|---|
| 1990 | 0-20 | 61.91 | 16.18 | 4.17 | 17.74 |
|  | 20-40 | 78.37 | 14.56 | 1.59 | 5.48 |
|  | 40-60 | 85.28 | 10.76 | 0.52 | 3.45 |
|  | 60-80 | 85.06 | 10.00 | 0.66 | 4.27 |
|  | 80-100 | 88.07 | 7.69 | 0.27 | 3.97 |
|  | All | 82.35 | 11.76 | 1.29 | 4.60 |
| 2000 | 0-20 | 62.17 | 18.37 | 5.36 | 14.11 |
|  | 20-40 | 77.77 | 14.95 | 2.38 | 4.91 |
|  | 40-60 | 83.19 | 12.00 | 0.46 | 4.35 |
|  | 60-80 | 86.21 | 9.13 | 0.46 | 4.20 |
|  | 80-100 | 87.00 | 8.03 | 0.28 | 4.68 |
|  | All | 83.16 | 11.91 | 1.40 | 3.54 |
| 2010 | 0-20 | 66.50 | 13.96 | 4.97 | 14.57 |
|  | 20-40 | 77.72 | 15.26 | 2.91 | 4.11 |
|  | 40-60 | 81.85 | 10.78 | 2.02 | 5.35 |
|  | 60-80 | 86.13 | 9.51 | 0.31 | 4.05 |
|  | 80-100 | 87.47 | 7.49 | 0.39 | 4.66 |
|  | All | 83.77 | 11.08 | 1.75 | 3.40 |

[1] Others include taxi, motorcycle, bicycle, walk, etc.

To explore the spatial patterns of commuting and neighborhood's mean wage rate, three bivariate choropleth maps are designed as shown in Figure 5. For each year, tracts are grouped into three categories of about equal frequency (i.e., 33, 66 and 100 percentile) in terms of mean commuting distance, and denoted by distinctive colors. In the meantime, as stated earlier, tracts are also grouped into five categories by mean wage rate percentile, and denoted by each color's darkness. The pattern of mean commuting distance shows a concentric pattern with increasing distance from the CBD for all three time periods, indicating the consistent significance of downtown Baton Rouge in influencing the commuting pattern. The pattern for mean wage rate displays a contrast between the southeast sector (higher wage) and the rest (including north and a narrow southwest strip with lower wage) in each year. While the general pattern for commuting distance remained concentric over years, the areas falling in each category changed. For example, the zones of short-distance commuting (i.e., 33 percentile marked in red) expanded to the southeast with medium to high income (in lighter red). This highlights that the areas with shorter-distance commuting in 1990 were mostly composed of low-wage earners, but expanded to areas with higher-wage earners. Tracts in the middle (40-60 percentile) wage group were mostly in the north in 1990 (mostly rural at the time), started to shift toward the middle area of Baton Rouge in 2000, and mostly were in the middle in 2010. This may help explain why this is the lone wage group with a trend of shortening commute distance from 1990 to 2010. Also note that areas of the lower wage (0-40 percentile) groups spread out from the central city area over time, and areas for higher wage (60-100 percentile) groups further stretched out from the south to further outskirts of development.

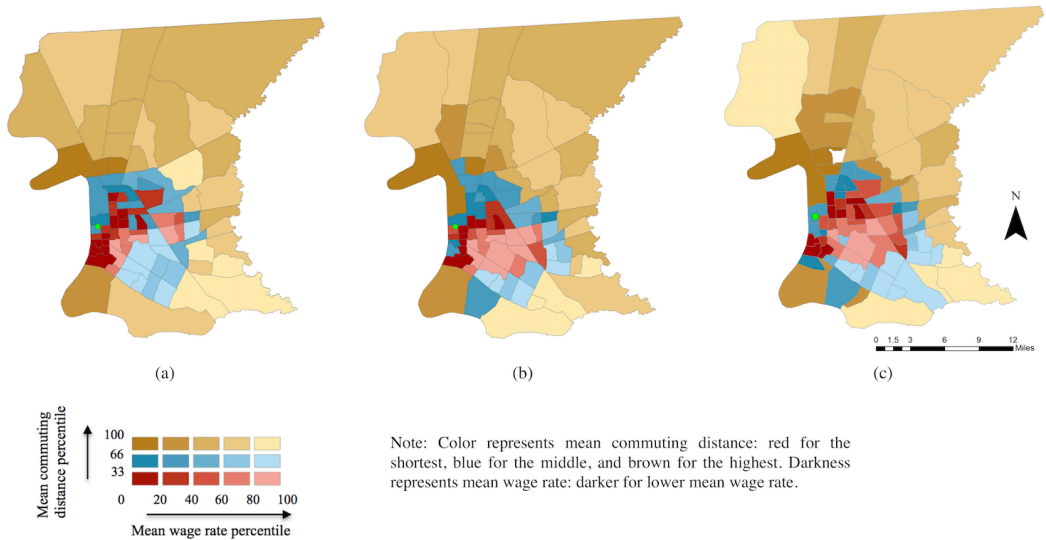

Figure 5. Mean commuting distance and mean wage rate: (a) 1990; (b) 2000; (c) 2010

### Commuting Patterns vs. Distribution of Wage Groups

The above section explores the effect of wage on commute by using the mean wage rate of tracts. This section analyzes the distribution of wage groups across census tracts in order to obtain more insight into the interaction between commuting and wage.

The CTPP data include the numbers of resident workers in various wage ranges. Different wage ranges were used among the CTPP 1990, 2000 and 2010. We divide workers in tracts into five wage groups (i.e., 0-15k, 15-35k, 35-50k, 50-75k, and 75k+), which is the only feasible classifications to be consistent over time. While all wage groups could be present in one tract with the same mean commute, their relative concentrations (e.g., percentages) vary. Limited by the aggregated data, we cannot single out which wage group(s) commute how much. However, if a certain group is consistently observed to be overrepresented in long-commuting areas and other groups are consistently found to concentrate in short-commuting areas, it is more likely than not that overall the former experiences longer commute than others (i.e., live further away from their jobs). Due to the ecological nature of the CTPP data, the inference from the analysis is merely suggestive and calls for validation by more in-depth analysis of individual data.

Given that workers of all wage groups could reside in the same tract, we formulate a null hypothesis for testing disparities in commuting lengths such as:

$H_0$ (null hypothesis): the ratios of a wage group in areas with above-average commute length are the same as those in areas with below-average commute length.

Here, the weighted average commute for the overall population in the study area is used as a benchmark for comparison. We are interested in examining whether a wage group is distributed disproportionately higher in areas of longer commute.

A conventional pooled t-test may be considered to test the above null hypothesis (Wang and Feliberty, 2010). For easy implementation for a large number of repetitive tests, we choose to use a weighted OLS regression to test the hypothesis (Ikram et al. 2015), where the tract population is used as the weight for appropriate adjustments in the error term. By doing so, a tract with more population is weighted heavier than one with less population. Such an adjustment is not feasible by the conventional pooled t-test.

Specifically, tracts in the study area are first split into two sets: Set 1 with below-average commute length are coded as "Flag = 0", and Set 2 are tracts with above-average commute length coded as "Flag = 1". Denoting the ratio of a wage group in a tract as $Y_w$, the regression model is defined as

$$Y_w = a + b \times Flag \qquad (3)$$

The intercept a is the average % of this wage group in Set 1 (i.e., tracts with below-average commute when Flag = 0). The average % of the wage group in Set 2 (i.e., tracts with above-average commute when Flag = 1) is simply reconstructed as a+b. The slope b is the difference between Sets 1 and 2, and its corresponding t value reveals the statistical significance for the difference.

Table 5. Wage groups in areas above or below the average commuting distance

|  | 1990 | | | 2000 | | | 2010 | | |
|---|---|---|---|---|---|---|---|---|---|
| Wage group | % with below average | % with above average | Difference in % | % with below average | % with above average | Difference in % | % with below average | % with above average | Difference in % |
|---|---|---|---|---|---|---|---|---|---|
| <15k | 50.3 | 38.5 | -11.8*** | 37.7 | 29.3 | -8.4** | 28.4 | 21.6 | -6.8* |
| 15-35k | 33.1 | 39.3 | 6.2*** | 32.9 | 36.9 | 4.0* | 30.9 | 30.3 | -0.6 |
| 35-50k | 9.0 | 12.7 | 3.7*** | 12.5 | 14.8 | 2.3* | 15.3 | 17.5 | 2.2 |
| 50-75k | 4.7 | 7.1 | 2.4** | 9.7 | 12.4 | 2.7* | 13.0 | 16.4 | 3.4* |
| >75k | 3.0 | 2.6 | -0.4 | 7.2 | 6.5 | -0.7 | 12.4 | 14.2 | 1.8 |

Note: *** indicates statistically significant at 0.001, ** significant at 0.01, * significant at 0.05.

Table 6. Wage groups in areas above or below the average commuting time

|  | 1990 | | | 2000 | | | 2010 | | |
|---|---|---|---|---|---|---|---|---|---|
| Wage group | % with below average | % with above average | Difference in % | % with below average | % with above average | Difference in % | % with below average | % with above average | Difference in % |
|---|---|---|---|---|---|---|---|---|---|
| <15k | 46.2 | 44.0 | -2.2 | 32.8 | 35.7 | 2.9 | 28.4 | 23.3 | -5.1 |
| 15-35k | 34.4 | 37.4 | 3.0 | 31.6 | 37.9 | 6.3*** | 28.4 | 32.9 | 4.5* |
| 35-50k | 10.1 | 11.3 | 1.2 | 14.0 | 12.9 | -1.1 | 15.6 | 16.7 | 1.1 |
| 50-75k | 5.8 | 5.6 | -0.2 | 12.0 | 9.4 | -2.6* | 13.4 | 15.0 | 1.6 |
| >75k | 3.6 | 1.8 | -1.8** | 9.6 | 4.0 | -5.6*** | 14.2 | 12.0 | -2.2 |

Note: *** indicates statistically significant at 0.001, ** significant at 0.01, * significant at 0.05

Table 5 and 6 report the significance test results of commuting distance and time, respectively. The results list the ratios of each wage group with below-average commute and above-average commute as well as their differences. For example, in 1990, 50.3% of lowest-wage workers (i.e., <15k) lived in areas with shorter commute distance than the overall population, while 38.5% lived in areas with above-average commute distance. The negative difference (i.e., -11.8) between both ratios is significant at 0.001 level, suggesting that workers in the lowest wage group in 1990 were significantly higher in areas with below-average commute distance, and thus enjoyed shorter commute distance in general. Similarly, we found that workers in the following three groups (i.e., 15-35k, 35-50k, and 50-75k) in 1990 were more concentrated in areas with above-average commute distance, indicating relatively longer commute lengths in these wage groups than the overall population. The highest-wage workers in 1990 appeared to have a slightly higher percentage living in areas with below-average commute distance, but not statistically significant. In a word, the above results demonstrate that workers in the lowest-wage group, in general, commuted significantly less than the overall population, while workers in the following three wage groups commuted significantly more than the overall population. However, no solid conclusions could be drawn for the highest-wage workers due to the insignificant results. Similar pattern was found in 2000 but not as clear as in 1990: only the lowest-wage group tended to concentrate in below-average commute distance areas with statistical significance, and concentrations of other groups in terms of areas of mean commute distance were not statistically significant. In 2010, the tendency of higher concentration of the lowest-wage group workers in areas of below-average commute distance remained significant, but the differences in concentration of other wage groups were not significant or minor (i.e., ratios of the 50-75k wage group in 2010 tended to be slightly higher in tracts of longer commute distance). In general, the convex shape pattern discovered in previous analysis is not observed in this analysis due to more insignificant results and different classifications of wage group. However, it does highlight the significant shorter commute distance for lowest-wage workers than the average in Baton Rouge 1990-2010, which might be largely determined by their limited mobility (e.g., low vehicle ownership, high dependency on bicycling, walking, or transit that is often feasible only in the central city area). Relevant policies such as adding more jobs near neighborhoods of low wage level or promoting affordable and accessible housing near job clusters may be considered for improvement.

In terms of commute time, there are several observations to make. First, the difference in concentration of the highest-wage workers was statistically significant in 1990 and 2000 (i.e., more in below-average commute time areas), but such a difference disappeared in 2010. Secondly, in 2000, workers in the 15-35k group were reported to disproportionately live in areas with above-average commute time (significant at 0.001 level). Again, no clear convex shape pattern was found in the association between commute time and wage; however, we do observe less commuting time with statistical significance for the highest-wage workers than the overall workers in 1990 and 2000. Given the ecological nature of the CTPP data, we refrain from further inference.

**Conclusions**

This research utilizes the CTPP data to analyze the commuting patterns (in both distance and time) and the association with income (wage as a surrogate) in Baton Rouge 1990-2010. In comparison to previous work, commuting distance is calibrated more accurately by Monte Carlo simulation of individual trips to mitigate aggregation error and scale effect.

The results indicate that commuting lengths vary across neighborhoods of different wage levels. The analysis based on neighborhoods mean wage rates demonstrates that higher-wage neighborhoods initially tended to commute further (in distance), but the trend was reversed toward less commuting in areas with the highest wage rates. The convex shape pattern is used to characterize neighborhoods in terms of the response of mean commuting distance to rising mean wage rate. The pattern of mean commute time was hardly consistent over time due to the variability of mode distributions across tracts of various mean wage rates. Given that a tract's mean wage rate cannot fully represent its real wage distribution pattern where workers of various wage rates reside, in addition, we examined the commuting patterns vs. distribution of wage groups across tracts as an analysis complementary to the above one. This analysis found no clear convex shape pattern in terms of both commute distance and time in 1990-2010; however, it highlighted the lowest-wage workers with shorter commute distance than the general workers (with statistical significance) as well as the highest-wage workers with less commute time (with statistical significance). The economic commute distance in the lowest-wage workers might be largely associated with their poor transport mobility (e.g., fewer location and transport options), while highest-wage workers are the groups who truly enjoy the efficient commute time. This finding may help target policy-makers on specific socio-demographic groups to improve the efficiency of planning and policies. For example, policy-makers could focus on strategies that increase vehicle ownership and public transportation opportunities for lowest-wage workers to help improve their limited mobility and thus promote social equity.

Methodologically, this research also contributes to the commuting studies with a land use-based simulation approach for more accurate commuting length measures. Indeed, data at detailed scales such as block level may have minor aggregation error. However, such detailed data are not always available in commuting studies due to reasons such as privacy. Regressions and statistical tests validate the advantage of our approach, which can be adopted in geospatial studies that are sensitive to spatial scales.

Using wage/income data at individual level would improve this research given that trips are derived at individual level by simulation; for example, we can build regression models at individual level rather than aggregated level (e.g., census tract). One major concern to this study is the ecological nature due to the limited data on wage (or income), which makes the above findings only suggestive but not necessarily applicable to individuals. Next steps in this research include evaluating the detected patterns in terms of commuting and wage using more detailed individual data on wage/income or detailed synthetic commuting flows. In addition, spatial statistics such as geographically weighted regression (GWR) could be adopted to study the spatial variation pattern between commute and wage when such data at finer scale (e.g., individual level) are available.

**Acknowledgements**

The Economic Development Assistantship (EDA) support from the Graduate School of Louisiana State University to Hu is gratefully acknowledged.